\documentclass[twocolumn]{revtex4}

\usepackage{amsmath,amssymb,graphicx,epsfig}
\usepackage{multirow,color}
\newcommand{\re}{\mathrm{Re}}
\newcommand{\im}{\mathrm{Im}}
\newcommand{\tr}{\mathrm{Tr}}

\begin{document}

\title{Entropy production in photovoltaic-thermoelectric nanodevices\\
 from the non-equilibrium Green's function formalism}

\author{Fabienne Michelini$^1$, Adeline Cr\'epieux$^2$, Katawoura Beltako$^1$}
\affiliation{$^1$Aix Marseille Univ,  Univ Toulon, CNRS, IM2NP, Marseille, France}
\affiliation{$^2$Aix Marseille Univ, Universit\'e de Toulon, CNRS, CPT, Marseille, France}

\begin{abstract}
We derive the expressions of photon energy and particle currents inside an open nanosystem interacting with light  using non-equilibrium Green's functions. The model allows different temperatures for the electron reservoirs, which basically defines a photovoltaic-thermoelectric hybrid. 
Thanks to these expressions, we formulate the steady-state entropy production rate to assess the efficiency of reversible photovoltaic-thermoelectric nanodevices. 
Next, quantum dot based nanojunctions are closely examined. We show that entropy production is always positive when one considers spontaneous emission of photons with a specific energy, while in general the emission spectrum is broadened, notably for strong coupling to reservoirs. In this latter case, when the emission is integrated over all the energies of the spectrum, we find that entropy production can reach negative values. This result provides matter to question the second law of thermodynamics for interacting nanosystems beyond the assumption of weak coupling.

\end{abstract}

\maketitle

\section{Introduction} 

At the end of the last century, structures grown or synthesized on a nanoscale have revealed new insights and potential applications which have  profoundly altered our vision of the technology. Indeed, nanostructures integrated into a device fundamentally change the behavior of electrons due to quantum effects like tunneling, confinement or entanglement.
Much hope has been focused on the dramatic potential of nanosciences and nanotechnologies. 
At the same time, these hopes have stimulated the use of many-body quantum methodologies to model electronic devices, like transistors~\cite{datta}, but also concepts of energy conversion like in photovoltaic~\cite{aeberhardJCEL11}, optomechanic~\cite{marquardt09} or thermoelectric~\cite{dubi11} devices. 

Among the methodologies of quantum statistics, non-equilibrium Green's function (NEGF) formalism is probably the most convenient one to deal with particle and energy transport in open interacting systems~\cite{datta}. In comparison, quantum master equation formalism is suited for regimes of weak coupling to electron reservoirs~\cite{roy15} or optically driven systems, while quantum cascade laser simulations have been carried out from NEGF formalism~\cite{wacker02}.
The community of quantum thermodynamics, and hence thermoelectricity, has developed experience in NEGF methodology~\cite{esposito15}, which has permitted to develop new insights in time-resolved conversion~\cite{crepieux11,dare16}, electron-electron interacting systems~\cite{zang15,azema12}, fundamental laws~\cite{yamamoto15,benenti15,whitney13}, and potential new paradigms~\cite{crepieux15,esposito15_prb}. 
For quantum optoelectronics and photovoltaics~\cite{henrickson02,aeberhard08,berbezier13APL,cavassilas15}, NEGF framework is more and more used, due to the fact that more and more applications involve electron transport at a nanoscale and quantum effects.

Novel technical directions are now being explored thanks to these powerful methodologies. 
One possible direction follows the idea of cogeneration, which combines outputted energy forms from a single sustainable energy source, like the simultaneous production of electrical and thermal energies from a single light source.
From a fundamental point of view, this direction is related to the thermodynamics of light~\cite{markvart16} which has naturally emerged in photovoltaics regarding the photon source as a thermal bath~\cite{wurfel}, and the thermoelectricity at contact between absorber and lead~\cite{humphrey05,rodiere15}. 
Closer to the commercial level, the creation and performance analysis of stacked photovoltaic-thermoelectric modules have been reported~\cite{park13,bjork15}. 
Incidentally, the promises of perovskites at the same time as photovoltaics~\cite{even15} and thermoelectrics~\cite{mettan15} also suggests that a combined energy conversion with these materials could be a success, using materials or nanostructures~\cite{ning15}.
Down to the nanoscale, theoretical designs based on nanostructures have been recently proposed for cooling~\cite{rey07,cleuren12,mari12,wang15} or joint cooling and electrical energy production from a photon source~\cite{entin15}. 
These proposals suggest to address the idea of cogeneration inside a unique module conceived at the nanoscale, which requires a deeper look at the energetic aspects of the light-matter coupling.

In this work, we derive the photon energy and particle currents in open nanosystems interacting with light using the framework of NEGFs. The Hamiltonian model is introduced in Sec.~\ref{sec:hamiltonian}, and the main lines of the derivation and results are given in Sec.~\ref{sec:currents}. This allows to calculate the entropy current flowing from the electron and photon reservoirs to the absorbing region of the device in Sec.~\ref{sec:entropycurrent}: we reshape and discuss the entropy production in terms of efficiencies for photovoltaic-thermoelectric nanodevices. 
Finally, we thoroughly examine a quantum-dot based architecture described with a two-level model in Sec.~\ref{sec:negentropy}: we show 
 that the entropy production is always positive at any coupling to electron reservoirs as long as one considers a unique photon energy for the emission process, but the entropy production can reach negative values if modifications are made on the model as it is traditionally done.


\section{Hamiltonian model}\label{sec:hamiltonian}
The Hamiltonian of a quantum nanosystem in contact with electronic reservoirs and interacting with light reads
\begin{equation}
H=H_0+H_T+H_{int}+H_L+H_R+H_{\gamma}\, ,
\end{equation}
where
\begin{subequations}
\begin{align}
H_0&=\sum_{n} \epsilon_{n}d_{n}^{\dagger}d _{n}\, ,\\
H_T&=\sum_{\alpha=L,R}\sum_{n\mathbf{k}} V_{\alpha n\mathbf{k}} {c}^{\dagger}_{\alpha n\mathbf{k}}{d}_{n}+\text{h.c.}\, ,\\
H_{int}&=\sum_{nm,\zeta \mathbf{q}} M_{nm,\zeta \mathbf{q}} d^\dagger_nd_mA_{\zeta \mathbf{q} }\label{Hint}\, ,\\
H_{\alpha \in \{L,R\}}&=\sum_{n\mathbf{k}} \epsilon_{\alpha n\mathbf{k}}{c}_{\alpha n\mathbf{k}}^{\dagger} {c}_{\alpha n\mathbf{k}}\, ,\\
H_{\gamma}&=\sum_{\zeta \mathbf{q}} \hbar \omega_{\zeta \mathbf{q} } \big[a_{\zeta \mathbf{q}}^{\dagger}a_{\zeta \mathbf{q}} + \frac{1}{2} \big]\, .
\end{align}
\end{subequations}
Subscript $0$ stands for the non-interacting and isolated nanosystem, $T$ for the transfer to the electron reservoirs, $int$ for the interaction with light, $\alpha$ for the the left ($L$) and right ($R$) electron reservoirs, and $\gamma$ for the photon bath. These expressions use the electron creation (annihilation) operators, $d_{n}^{\dagger}$ ($d_{n}$) in the central region and ${c}^{\dagger}_{\alpha n\mathbf{k}}$ (${c}_{\alpha n\mathbf{k}}$) in the electron reservoirs. On the other side, $a_{\zeta \mathbf{q}}^{\dagger}$ ($a_{\zeta \mathbf{q}}$) and $A_{\zeta \mathbf{q} }=a_{\zeta \mathbf{q} }+a_{\zeta -\mathbf{q} }^\dagger$ are photon operators of the photon bath, with $ \mathbf{q}$ the wave vector of the radiation, and $\zeta $ one of the two directions of polarization perpendicular to the propagation. The interacting central region is coupled to the electron reservoirs via parameters $ V_{\alpha n\mathbf{k}}$ while it is coupled to the light radiation via parameters $M_{nm,\zeta \mathbf{q}}$.
Usually, calculation for optoelectronics relies on the dipole approximation: $ \mathbf{q}\cdot \mathbf{r}\ll 1$ where $\mathbf{r}$ is the spatial coordinate~\cite{mandelwolf}.

We introduce the notations used in this paper: 
$I^E_\beta=-\langle \dot H_{\beta } \rangle$ the energy current, 
$I_\beta= -\langle \dot N_{\beta } \rangle$ with $N_{\beta \in \{L,R\}}=\sum_{n\mathbf{k}} {c}_{\beta n\mathbf{k}}^{\dagger} {c}_{\beta n\mathbf{k}}$ and $N_{\gamma}=\sum_{\zeta \mathbf{q}} a_{\zeta \mathbf{q}}^{\dagger}a_{\zeta \mathbf{q}}$ the particle current, 
and finally $I^h_\beta = -\langle \dot H_{\beta } \rangle +\mu_{\beta} \langle \dot N_{\beta } \rangle=I^E_\beta-\mu_{\beta}I^p_\beta$  for the heat current~\cite{crepieux11}, where $\mu_\beta$ is the (electro)chemical potential of reservoir $\beta$. In the case of electron reservoirs, we will also use $I^e_{\beta \in {L,R}}$ for the electrical current.
All currents are flowing from the $\beta \in L,R,\gamma$ reservoir to the central region.


\section{Photon currents}\label{sec:currents}

\subsection{Photon energy current}\label{photonenergycurrent}

We derived the formal expression of the photon energy current $I^E_{\gamma} (t)=- \langle \dot H_{\gamma} \rangle (t)$ inside an optoelectronic device following the first order Born approximation within the Keldysh's formalism \cite{haug,mahan}. 
From the Heisenberg equation $ \dot H_{\gamma} = -\frac{i}{\hbar}\left[H_{\gamma},H\right]$, the energy current can be expressed in terms of expectation values on mixed operators which combine electron and photon operators 
\begin{eqnarray}\label{eq:formalIEphoton}
I^E_{\gamma}  (t) &=&  \frac{1}{\hbar}  \re \sum_{\zeta \mathbf{q}} \hbar\omega_{\zeta \mathbf{q} } \mathbf{M}_{\zeta \mathbf{q}}\mathbf{\mathfrak{B}}^<_{\zeta \mathbf{q}}(t,t)\, ,
\end{eqnarray}
where ${ \mathfrak{B}^<_{mn,\zeta \mathbf{q}}(t,t)}=i\langle d^{\dag}_n (t) d_m(t) B_{\zeta \mathbf{q}}(t) \rangle$ with $B_{\zeta \mathbf{q}}(t)=a_{\zeta \mathbf{q} }( t) - a^\dagger_{\zeta -\mathbf{q}}(t)$. Here we use matrix forms to encode level and/or eventual space-discretization indices. 
In the framework of the Keldysh formalism, we sought the expression of the contour ordered mixed Green's function 
\begin{eqnarray}\label{mixedfunction}
{ \mathfrak{B}^t_{mn,\zeta \mathbf{q}}(\tau,\tau')}&=& -i\langle T d_m(\tau) d^{\dag}_n (\tau') B_{\zeta \mathbf{q}}(\tau) \rangle~,
\end{eqnarray}
where $T$ is the time-ordering operator.
The main lines of the derivation follow the first order Born approximation~\cite{mahan}, which consists in switching to the interaction picture, developing the time evolution operator up to the second order in the electron-boson interaction parameter, using Wick's theorem, verifying the cancellation of the disconnected graphs, including higher order contributions with self-consistency and finally performing the Langreth's rules for the analytic continuation.
We thus obtain
\begin{eqnarray}
\sum_{\zeta \mathbf{q}} \hbar\omega_{\zeta \mathbf{q} } \mathbf{M}_{\zeta \mathbf{q}} \mathbf{\mathfrak{B}}^<_{\zeta \mathbf{q}}(\tau,\tau') &=& \tr  \int d\tau_1 \mathbf{\Xi}^t_{\gamma} (\tau,\tau_1) \nonumber \\
&&\times \mathbf{G}^t(\tau_1,\tau') \, ,
\end{eqnarray}
with
\begin{eqnarray}
{\bf \Xi}^t_{\gamma}(\tau,\tau_1)&=& \sum_{\zeta \mathbf{q}}  \hbar\omega_{\zeta \mathbf{q} }   \mathbf{M}_{\zeta \mathbf{q}} \mathbf{G}^t(\tau,\tau_1)\widetilde D_{\zeta \mathbf{q}}^{0t}(\tau,\tau_1) \mathbf{M}_{\zeta \mathbf{q}}\, . \nonumber \\
\end{eqnarray}
The expression of ${ \mathfrak{B}^<_{mn,\zeta \mathbf{q}}(t,t)}$ is then deduced from the Langreth's rules, which finally provides the photon energy current from Eq.~(\ref{eq:formalIEphoton}),
\begin{eqnarray}
I^E_{\gamma}  (t)  &=&\frac{1}{\hbar}\re \mathrm{Tr} \int dt_1 \big[{\bf G}^r(t,t_1){\bf \Xi}^<_{\gamma}(t_1,t)\nonumber \\
&&+{\bf G}^<(t,t_1){\bf  \Xi}^a_{\gamma}(t_1,t)\big]\, .
\end{eqnarray}
These expressions use the standard Green's functions for the electrons inside the central region $\mathbf{G}^t(\tau,\tau')$, defined as $G^t_{nm}(\tau,\tau')=-i \langle T d_n (\tau)d^{\dag}_m (\tau')\rangle $. On the other side, we introduce the  photon Green's function $\widetilde D_{\zeta \mathbf{q}}^{0t}(\tau,\tau_1)=-i \langle TB_{\zeta \mathbf{q}}(\tau) A_{\zeta \mathbf{q}}(\tau_1)\rangle$. Function  $\widetilde D_{\zeta \mathbf{q}}^{0t}(\tau,\tau_1)$ differs in a negative sign from $D_{\zeta \mathbf{q}}^{0t}(\tau,\tau_1)= -i  \langle TA_{\zeta \mathbf{q}}(t) A_{\zeta \mathbf{q}}(\tau_1)\rangle$ which appears in the derivation of the Dyson's equation for an electron interacting with bosons~\cite{mahan}.

For steady-state devices, we obtain
\begin{eqnarray} \label{IEphoton}
 I^E_{\gamma} &=&\frac{1}{\hbar}\re \mathrm{Tr} \int \frac{d\epsilon}{2\pi} \big[{\bf G}^r{\bf \Xi}^<_{\gamma}+{\bf G}^<{\bf  \Xi}^a_{\gamma}\big](\epsilon)\, ,
\end{eqnarray}
where
\begin{eqnarray}\label{xi_tot}
{\bf \Xi}^{\lessgtr}_{\gamma}(\epsilon)&=& \sum_{\zeta \mathbf{q}}   \hbar \omega _{\zeta \mathbf{q}} \mathbf{M}_{\zeta \mathbf{q}} \big[\pm \underbrace{ N_{\zeta \mathbf{q}} \mathbf{G}^\lessgtr(\epsilon\mp \hbar \omega _{\zeta \mathbf{q}} )}_{\text{absorption contribution}} \nonumber \\
&  \mp  &\underbrace{ (N_{\zeta \mathbf{q}} \overbrace{+1}^{\text{spontaneous}}) \mathbf{G}^\lessgtr(\epsilon\pm \hbar \omega _{\zeta \mathbf{q}} )}_{\text{emission contribution}}\big] \mathbf{M}_{\zeta \mathbf{q}} \, ,
\end{eqnarray}
and
\begin{eqnarray}
{\bf \Xi}^{r,a}_{\gamma}(\epsilon)&=&\pm\frac{1}{2}\big[{\bf \Xi}^{>}_{\gamma}(\epsilon)-{\bf \Xi}^{<}_{\gamma}(\epsilon)\big] \nonumber \\
&&-i{\cal P} \int d\epsilon' \frac{{\bf \Xi}^{>}_{\gamma}(\epsilon')-{\bf \Xi}^{<}_{\gamma}(\epsilon')}{\epsilon-\epsilon'} \, ,
\end{eqnarray}
$N_{\zeta \mathbf{q}}$ is the occupation number of the radiation modes $\zeta \mathbf{q}$. These modes form the photon bath which is assumed to be in an equilibrium or quasi-equilibrium state of temperature $T_{\gamma}$ and chemical potential $\mu_\gamma$.

In Eq.~(\ref{xi_tot}), we outline the three essential contributions for the radiative processes of the electron-photon interaction~\cite{mandelwolf}: the two induced processes which include the absorption ($abs$) and the stimulated emission ($em,st$), and the spontaneous emission ($em,sp$) which is independent of the occupation number of the photon bath and is non-zero in the vacuum state. 
The photon energy current can thus be split according to different viewpoints
\begin{eqnarray}
I^E_{\gamma} &=&I^E_{abs}+I^E_{em,st}+I^E_{em,sp}\\
&=&I^E_{ind}+I^E_{em,sp} \\
&=&I^E_{abs}+I^E_{em} \, ,
\end{eqnarray}
as the function ${\bf \Xi}_{\gamma}(\epsilon)$ which has the dimension of $energy^2$
\begin{eqnarray}
{\bf \Xi}^{\lessgtr}_{\gamma} &=&{\bf \Xi}^{\lessgtr}_{abs}+{\bf \Xi}^{\lessgtr}_{em,st}+{\bf \Xi}^{\lessgtr}_{em,sp}~,
\end{eqnarray}
with
\begin{subequations} \begin{align}
{\bf \Xi}^{\lessgtr}_{abs}(\epsilon)&=\pm  \sum_{\zeta \mathbf{q}}   \hbar\omega_{\zeta \mathbf{q} } \mathbf{M}_{\zeta \mathbf{q}} N_{\zeta \mathbf{q}} \mathbf{G}^\lessgtr(\epsilon\mp \hbar \omega _{\zeta \mathbf{q}} ) \mathbf{M}_{\zeta \mathbf{q}}\label{xi_a}  \, ,\\
{\bf \Xi}^{\lessgtr}_{em,st}(\epsilon)&= \mp \sum_{\zeta \mathbf{q}}   \hbar\omega_{\zeta \mathbf{q} } \mathbf{M}_{\zeta \mathbf{q}} N_{\zeta \mathbf{q}} \mathbf{G}^\lessgtr(\epsilon\pm \hbar \omega _{\zeta \mathbf{q}} ) \mathbf{M}_{\zeta \mathbf{q}} \label{xi_st}  \, ,\\
{\bf \Xi}^{\lessgtr}_{em,sp}(\epsilon)&= \mp \sum_{\zeta \mathbf{q}}   \hbar\omega_{\zeta \mathbf{q} } \mathbf{M}_{\zeta \mathbf{q}} \mathbf{G}^\lessgtr(\epsilon\pm \hbar \omega _{\zeta \mathbf{q}} ) \mathbf{M}_{\zeta \mathbf{q}}  \, .\label{xi_sp}
\end{align} 
\end{subequations}

\subsection{Energy conservation}
\label{energyconservation}

In the case of the self-consistent Born approximation, the energy has to be conserved in the total system~\cite{baym}
\begin{equation}
\langle \dot H \rangle=\langle \dot H_L+ \dot H_R+ \dot H_{\gamma} + \dot H_0 +\dot H_T + \dot H_{int}\rangle=0\,.
\end{equation}
The two first terms are known from Ref.~\onlinecite{crepieux11}, $\dot H_{\alpha \in L,R}=- \frac{2}{h} \re \mathrm{Tr} \int d\epsilon \epsilon \big[{\bf G}^r{\bf \Sigma}^<_{\alpha}+{\bf G}^<{\bf  \Sigma }^a_{\alpha}\big] (\epsilon)$ with ${\bf \Sigma}^{<,a}_{\alpha}$ the reservoir self-energies \cite{haug}. 
Energy currents related to the central region, the transfer process and the light-matter coupling are zero for steady-state operating: $ \langle \dot H_0 \rangle=0$, $\langle \dot H_T \rangle =0$, and $ \langle \dot H_{int} \rangle=0$.
We verified the energy conservation requirement starting with expressions~(\ref{IEphoton}) together with $\langle \dot H_L \rangle $ and $\langle \dot H_R \rangle $ from Ref.~\onlinecite{crepieux11}.
For the calculations, we used the property $\tr  \Big[ {\bf \Sigma}^<\mathbf{G}^>-  {\bf \Sigma}^>\mathbf{G}^< \Big]=0$ with $  {\bf \Sigma}^{\lessgtr}=\sum_{\alpha \in L,R,int} {\bf \Sigma}^{\lessgtr}_{\alpha}$ the total self-energy~\cite{haug}, in order to eliminate ${\bf  \Sigma }^{\lessgtr}_{L,R}$. Then, we evidenced the following quantity $\epsilon {\bf \Sigma}^{\lessgtr}_{\gamma}(\epsilon)-\hbar \omega_{\zeta \mathbf{q}} {\bf \widetilde \Sigma}^{\lessgtr}_{\gamma}(\epsilon)$, in order to perform change of integration variables of type $\epsilon'=\epsilon \pm \hbar \omega_{\zeta \mathbf{q}}$.

\subsection{Absorption and emission rates}\label{particulecurrent}

Similarly, the photon current $I_{\gamma} (t)= -\langle \dot N_{\gamma} \rangle (t) $ can be derived relying on previous mixed Green's functions $\mathfrak{B}^t_{mn,\zeta \mathbf{q}}(t,t')$ defined Eq.~(\ref{mixedfunction}). We get
\begin{eqnarray}
I_{\gamma} (t)&=&\frac{1}{\hbar}\re \mathrm{Tr} \int dt_1 \big[{\bf G}^r(t,t_1){\bf \widetilde \Sigma }^<_{\gamma}(t_1,t)\nonumber\\
&&+{\bf G}^<(t,t_1){\bf  \widetilde \Sigma}^a_{\gamma}(t_1,t)\big] \, ,
\end{eqnarray}
with
\begin{eqnarray}
{\bf \widetilde \Sigma}^{\lessgtr}_{\gamma}(t,t')&=&\sum_{\zeta \mathbf{q}}  \mathbf{M}_{\zeta \mathbf{q}}(t)  \mathbf{G}^\lessgtr(t,t')\widetilde D_{\zeta \mathbf{q}}^{0\lessgtr}(t,t')  \mathbf{M}_{\zeta \mathbf{q}} (t') \, . \nonumber \\
\end{eqnarray}

For steady-state operation, we obtained
\begin{eqnarray}\label{Ipphoton}
I_{\gamma} &=&\frac{1}{\hbar}\re \mathrm{Tr} \int \frac{d\epsilon}{2\pi} \big[{\bf G}^r{\bf \widetilde \Sigma}^<_{\gamma}+{\bf G}^<{\bf  \widetilde \Sigma}^a_{\gamma}\big](\epsilon)  \, ,\nonumber\\
\end{eqnarray}
where
\begin{eqnarray}
{\bf \widetilde \Sigma}^{\lessgtr}_{\gamma}&=& \sum_{\zeta \mathbf{q}}  \mathbf{M}_{\zeta \mathbf{q}} \big[ \pm N_{\zeta \mathbf{q}} \mathbf{G}^\lessgtr(\epsilon\mp \hbar \omega _{\zeta \mathbf{q}} )\nonumber\\
&&\mp (N_{\zeta \mathbf{q}} +1) \mathbf{G}^\lessgtr(\epsilon\pm \hbar \omega _{\zeta \mathbf{q}} )\big]  \mathbf{M}_{\zeta \mathbf{q}}  \, ,
\end{eqnarray}
and still
\begin{eqnarray}
{\bf \widetilde \Sigma}^{r,a}_{\gamma}(\epsilon)&=&\pm\frac{1}{2}\big[{\bf \widetilde \Sigma}^{>}_{\gamma}(\epsilon)-{\bf \widetilde \Sigma}^{<}_{\gamma}(\epsilon)\big] \nonumber \\
&&-i{\cal P} \int d\epsilon' \frac{{\bf \widetilde \Sigma}^{>}_{\gamma}(\epsilon')-{\bf \widetilde \Sigma}^{<}_{\gamma}(\epsilon')}{\epsilon-\epsilon'} \, .
\end{eqnarray}
It is worth comparing the function ${\bf \widetilde \Sigma}_{\gamma}$ with the interaction self-energy ${\bf \Sigma}_{\gamma}$ which happens in the Dyson equation for an electron interacting with the light radiation~\cite{mahan,aeberhard08}
\begin{eqnarray}
{\bf \Sigma}^{\lessgtr}_{\gamma}(\epsilon)&=& \sum_{\zeta \mathbf{q}}  \mathbf{M}_{\zeta \mathbf{q}} \big[ N_{\zeta \mathbf{q}} \mathbf{G}^\lessgtr(\epsilon\mp \hbar \omega _{\zeta \mathbf{q}} )\nonumber\\
&&+ (N_{\zeta \mathbf{q}} +1) \mathbf{G}^\lessgtr(\epsilon\pm \hbar \omega _{\zeta \mathbf{q}} )\big]  \mathbf{M}_{\zeta \mathbf{q}}   \, .
\end{eqnarray}
Sign changes between ${\bf \widetilde \Sigma}^<_{\gamma}$ and ${\bf \Sigma}^<_{\gamma}$ are intuitive: absorption(emission) means that a photon is flowing from the photon bath(central region) to the central region(photon bath). Without these sign changes, $\re \mathrm{Tr} \int \frac{d\epsilon}{2\pi} \big[{\bf G}^r{\bf \Sigma}^<_{\gamma}+{\bf G}^<{\bf \Sigma}^a_{\gamma}\big](\epsilon) =0$ (to be compared with Eq.~(\ref{Ipphoton})), which fulfills the condition of current conservation along the nanodevice~\cite{haug}. 

Similarly to the case of the photon energy current, it is meaningful to distinguish between the three radiative processes of the electron-photon interaction throughout
\begin{eqnarray}\label{processes}
{\bf \widetilde \Sigma}^{\lessgtr}_{\gamma}&=& \underbrace{{\bf \widetilde \Sigma}^{\lessgtr}_{abs}+{\bf \widetilde \Sigma}^{\lessgtr}_{em,st}}_{ind\text{uced processes}}+{\bf \widetilde \Sigma}^{\lessgtr}_{em,sp}\, ,
\end{eqnarray}
with
\begin{subequations} \begin{align}
 {\bf \widetilde \Sigma}^{\lessgtr}_{abs}(\epsilon)=&\pm  \sum_{\zeta \mathbf{q}} N_{\zeta \mathbf{q}}  \mathbf{M}_{\zeta \mathbf{q}} \mathbf{G}^\lessgtr(\epsilon\mp \hbar \omega _{\zeta \mathbf{q}} ) \mathbf{M}_{\zeta \mathbf{q}} \, ,\\
 {\bf \widetilde \Sigma}^{\lessgtr}_{em,st}(\epsilon)=&\mp \sum_{\zeta \mathbf{q}} N_{\zeta \mathbf{q}}\mathbf{M}_{\zeta \mathbf{q}}  \mathbf{G}^\lessgtr(\epsilon\pm \hbar \omega _{\zeta \mathbf{q}} ) \mathbf{M}_{\zeta \mathbf{q}}  \, , \\
 {\bf \widetilde \Sigma}^{\lessgtr}_{em,sp}(\epsilon)=&\mp \sum_{\zeta \mathbf{q}} \mathbf{M}_{\zeta \mathbf{q}} \mathbf{G}^\lessgtr(\epsilon\pm \hbar \omega _{\zeta \mathbf{q}} ) \mathbf{M}_{\zeta \mathbf{q}}  \,.
\end{align} 
\end{subequations}

The derivation of  $I_{\gamma} $ Eq.~(\ref{Ipphoton}) provides general expressions for the radiative rates in the stationary case. Indeed, we decompose $\langle \dot N_{\gamma} \rangle = -R_{abs}+R_{em,st} +R_{em,sp}$ and then identify (using the cycling property of the trace)
\begin{subequations}\label{rates} 
\begin{align}
R_{abs} =& \frac{1 }{h}  \sum_{\zeta \mathbf{q}}  N_{\zeta \mathbf{q}} \int d\epsilon  \mathcal{T}^{-}(\epsilon,\hbar \omega_{\zeta \mathbf{q}} ) \label{rate_a} \, ,\\
R_{em,st} =&\frac{1}{h}  \sum_{\zeta \mathbf{q}} N_{\zeta \mathbf{q}} \int d\epsilon  \mathcal{T}^{+}(\epsilon,\hbar \omega_{\zeta \mathbf{q}} ) \label{rate_emst}  \, , \\
R_{em,sp} =&\frac{1}{h}  \sum_{\zeta \mathbf{q}}  \int d\epsilon \mathcal{T}^{+}(\epsilon,\hbar \omega_{\zeta \mathbf{q}} ) \label{rate_emsp}  \, ,
\end{align} 
\end{subequations}
where
\begin{eqnarray}\label{traces}
\mathcal{T}^{\pm}(\epsilon,\hbar \omega_{\zeta \mathbf{q}} )&=&  \tr \big[ \mathbf{M}_{\zeta \mathbf{q}} \mathbf{G}^<(\epsilon \pm \hbar \omega_{\zeta \mathbf{q}} )  \mathbf{M}_{\zeta \mathbf{q}}{\bf G}^>(\epsilon)  \big] \, . \nonumber \\
\end{eqnarray}
Expressions (\ref{rate_a}-\ref{rate_emsp}) reiterate the formula provided by Aeberhard in Ref.~\onlinecite{aeberhard11} from analogy between the Boltzmann and Dyson equations. 

\subsection{Spectral photon currents}
\label{spectralcurrents}

From decomposition Eq.~(\ref{processes}), it is possible to derive the three spectral photon currents using the photon density of states $\mathcal{D}_{\zeta \mathbf{u}}(\hbar \omega )$
\begin{subequations} 
\begin{align}
r^{abs}_{\zeta \mathbf{u}} (\hbar \omega ) =& N_{\zeta \mathbf{u}} (\hbar \omega )\mathcal{D}_{\zeta \mathbf{u}} (\hbar \omega)\frac{1}{h} \int d\epsilon  \mathcal{T}^{-}_{\zeta \mathbf{u}} (\epsilon,\hbar \omega )  \, ,\\
r^{em,st}_{\zeta \mathbf{u}}  (\hbar \omega  ) =& -N_{\zeta \mathbf{u}}(\hbar \omega ) \mathcal{D}_{\zeta \mathbf{u}} (\hbar \omega  )\frac{1}{h} \int d\epsilon\mathcal{T}^{+}_{\zeta \mathbf{u}} (\epsilon,\hbar \omega ) \, , \\
r^{em,sp}_{\zeta \mathbf{u}}  (\hbar \omega )  =&  -\mathcal{D}_{\zeta \mathbf{u}} (\hbar \omega )\frac{1}{h} \int d\epsilon  \mathcal{T}^{+}_{\zeta \mathbf{u}} (\epsilon,\hbar \omega)   \label{luminance} \, .
\end{align} 
\end{subequations}
We have introduced the direction of light propagation $\mathbf{u}=\mathbf{q}/||\mathbf{q}||$ and abbreviated the frequency by writing $\omega$.

Within theses derivations, the radiation is treated as a third terminal, in contrast with other developments where the photon Green's functions are fully taken into accounts with their own dynamics~\cite{richter08}. However, these derivations allow us to provide radiation properties from the knowledge of the matter, in terms of electron Green's functions, via the trace of $\mathbf{M}_{\zeta \mathbf{u}}  \mathbf{G}^<(\epsilon \pm \hbar \omega ) \mathbf{M}_{\zeta \mathbf{u}} {\bf G}^>(\epsilon)$ (see Eq.~(\ref{traces})).  This function depends on both the electron and photon energies, and it is connected to the polarization insertion of the interaction dynamics~\cite{fetterwalecka}.
It is also interesting to introduce the induced spectral current $r^{ind}=r^{abs} +r^{em,st} $ given by
\begin{eqnarray}\label{eq:rind}
r^{ind}_{\zeta \mathbf{u}}  (\hbar \omega ) &= & N_{\zeta \mathbf{u}} (\hbar \omega )   \mathcal{D}_{\zeta \mathbf{u}}(\hbar \omega  ) \mathcal{A}_{\zeta \mathbf{u}} (\hbar \omega ) \, ,
\end{eqnarray}
where 
\begin{equation}\label{eq:A}
\mathcal{A} _{\zeta \mathbf{u}} (\hbar \omega ) = \frac{1}{h} \int d\epsilon  \big[  \mathcal{T}^{-}_{\zeta \mathbf{u}} - \mathcal{T}^{+}_{\zeta \mathbf{u}} \big](\epsilon,\hbar \omega )
\end{equation}
is a rate of net absorption (if $\mathcal{A} >0$) or gain (if $\mathcal{A} <0$) in the optoelectronic device. 
Taking advantage of the equality of $\int d\epsilon \mathcal{T}^{-}_{\zeta \mathbf{u}}(\epsilon,\hbar \omega)=\int d\epsilon \mathcal{T}^{-}_{\zeta \mathbf{u}}(\epsilon+\hbar \omega,\hbar \omega)$, the spectral current $r^{ind}$ is shaped into
\begin{eqnarray}
r^{ind}_{\zeta \mathbf{u}} (\hbar \omega ) &= & \frac{1}{h}\mathcal{D}_{\zeta \mathbf{u}}(\hbar \omega) N_{\zeta \mathbf{u}} (\hbar \omega)  \int d\epsilon  \mathcal{T}^{+}_{\zeta \mathbf{u}} (\epsilon,\hbar \omega  )  \nonumber \\
&& \times {\cal B}^{-1}_{\zeta \mathbf{u}}(\epsilon,\hbar \omega)  \, ,
\end{eqnarray}
with%
\begin{eqnarray}\label{Bintrace}
{\cal B} _{\zeta \mathbf{u}}(\epsilon,\hbar \omega )  &= & \Bigg[\frac{   \tr \big[ \mathbf{M} _{\zeta \mathbf{u}}  \mathbf{G}^<(\epsilon) \mathbf{M}_{\zeta \mathbf{u}} {\bf G}^>(\epsilon+ \hbar \omega) \big]} {  \tr \big[ \mathbf{M}_{\zeta \mathbf{u}}  \mathbf{G}^<(\epsilon + \hbar \omega) \mathbf{M}_{\zeta \mathbf{u}} {\bf G}^>(\epsilon) \big]} -1\Bigg]^{-1}  \nonumber \\
&=& \Bigg[\frac{\mathcal{T}^{-}_{\zeta \mathbf{u}}(\epsilon+\hbar \omega,\hbar \omega)} { \mathcal{T}^{+}_{\zeta \mathbf{u}}(\epsilon,\hbar \omega )} -1\Bigg]^{-1}  \, .
\end{eqnarray}

Using the dimensionless function $ {\cal B} _{\zeta \mathbf{u}}(\epsilon,\hbar \omega ) $, we finally formulate the photon particle and energy currents as follows
\begin{eqnarray}\label{eq:ne_photoncurrent}
I_{\gamma}&= &\frac{1}{h} \sum_{\zeta} \int d\Omega_{\mathbf{u}} \int d(\hbar \omega) \mathcal{D} _{\zeta \mathbf{u}} (\hbar \omega)  \nonumber \\
&& \times  \int d\epsilon  \mathcal{T}^{+} _{\zeta \mathbf{u}}(\epsilon,\hbar \omega )  \Big[ N_{\zeta \mathbf{u}}(\hbar \omega)  {\cal B}^{-1} _{\zeta \mathbf{u}}(\epsilon,\hbar \omega ) -1 \Big]  ~,\nonumber \\~ \\
\label{eq:ne_photonenergycurrent}
I^E_{\gamma}&= &\frac{1}{h} \sum_{\zeta} \int d\Omega_{\mathbf{u}} \int d(\hbar \omega) \mathcal{D} _{\zeta \mathbf{u}}(\hbar \omega )\hbar \omega \nonumber \\
&& \times  \int d\epsilon  \mathcal{T}^{+} _{\zeta \mathbf{u}}(\epsilon,\hbar \omega)  \Big[ N_{\zeta \mathbf{u}}(\hbar \omega)  {\cal B}^{-1} _{\zeta \mathbf{u}}(\epsilon,\hbar \omega ) -1 \Big] ~,\nonumber \\
\end{eqnarray}
where $d\Omega_{\mathbf{u}}$ is the elementary solid angle around the direction ${\mathbf{u}}$ along which the light propagates.
It is interesting to point out the similar expressions we have for these currents: they are both written as the product of a two-dimensional spectral quantity $\mathcal{D}_{\zeta \mathbf{u}}(\hbar \omega  ) \mathcal{T}^{+}_{\zeta \mathbf{u}} (\epsilon,\hbar \omega )  \Big[ N_{\zeta \mathbf{u}}(\hbar \omega  )   {\cal B}^{-1}_{\zeta \mathbf{u}} (\epsilon,\hbar \omega) -1 \Big] $ multiplied by the photon energy $\hbar \omega$ at the power zero for the particle current, and at the power one for the energy current.

\subsection{Quasi-equilibrium limits}
Within NEGF formalism, the electron-photon interaction is described using the self-consistent Born approximation in terms of electron and photon Green's functions. The approach is original in the sense that it is in fact not necessary to define local thermodynamic parameters to obtain particle, energy or entropy currents which flow outside the out-of-equilibrium central region. In devices where the central region reaches the nanoscale, particles experience non-thermal states while the device is working. It is not a simple task to define local temperature, and electrochemical potential in the interacting central region~\cite{whitneyPE16,meair14}. Indeed, all the particle statistics is encoded in NEGF formalism~\cite{bruus}.
However, if NEGFs can be represented by quasi-equilibrium Green's functions, they will verify a \textit{Kubo-Martin-Schwinger} relation~\cite{balzer} $
 \mathbf{G}^>(\epsilon)= \mathbf{G}^<(\epsilon)e^{\frac{\epsilon-\mu}{k_B T}}$, where $\mu$ and $T$ represent the electronic chemical potential and temperature respectively.
This relation generalizes the following properties of the Fermi and Bose functions: $1-f(\epsilon)=e^{\frac{\epsilon-\mu}{k_B T}} f(\epsilon)$ and $N(\hbar \omega) +1=e^{\frac{\hbar \omega}{k_B T}} N(\hbar \omega)$.

More generally, let us consider the case of a semiconductor in which electrons inside the conduction band,  and holes inside the valence band experience separate quasi-equilibrium states characterized by two different chemical potentials and temperatures, $\mu_{c,v}$ and $T_{c,v}$. In that case, the diagonal components of Green's functions follow local Kubo-Martin-Schwinger relations
\begin{equation}
\big[  \mathbf{G} ^>\big]_n(\epsilon)=\big[   \mathbf{G}^<\big]_n(\epsilon)e^{\frac{\epsilon-\mu_n}{k_B T_n}}~,
\end{equation}
where $(n \in c,v)$ refers to the band index.
Thanks to these relations, simplifications occur in the expression of ${\cal B}$, Eq.~(\ref{Bintrace}). In particular for $T=T_c=T_v$, ${\cal B}$ no longer depends on the electron energy $\epsilon$, and it follows
\begin{equation}
 {\cal B} (\hbar \omega )=\Bigg[\exp \Big(\frac{ \hbar \omega -(\mu_c-\mu_v)}{k_B T}\Big) -1\Bigg]^{-1}~,
\end{equation}
in which one can define
$T_E=T$ and $\mu_E=(\mu_c-\mu_v)$, the temperature and chemical potential of the spontaneously emitted radiation~\cite{wurfel82}.
Hence we obtain the full Bose-Einstein statistic function that happens in the so-called generalized Planck's law for the emission~\cite{lasher64,wurfel82}, that was also discussed in photovoltaic cells of quantum dot arrays using NEGFs~\cite{berbezier15}, and notably used to determine the thermopower from optical measurements~\cite{gibelli16}.
 
In the quasi-equilibrium limit, ${\cal B}$ does not depend on $\epsilon$, which allows us to write the spectral emission current as
\begin{eqnarray}
r^{em,sp}_{\zeta \mathbf{u}} (\hbar \omega  )  &=&- {\cal B}(\hbar \omega)  \mathcal{D}_{\zeta \mathbf{u}}(\hbar \omega) \mathcal{A}_{\zeta \mathbf{u}}(\hbar \omega) ~.
\end{eqnarray}
Using Eqs.~(\ref{eq:rind}) and (\ref{eq:A}), the two quasi-equilibrium limits of the photon particle and energy currents finally read as
\begin{eqnarray}\label{eq:equilibrium_photoncurrent}
I_{\gamma}&= & \sum_{\zeta} \int d\Omega_{\mathbf{u}} \int d(\hbar \omega)  \mathcal{D}_{\zeta \mathbf{u}}(\hbar \omega ) \nonumber \\
&& \times \big[N(\hbar \omega) -  {\cal B}(\hbar \omega) \big] \mathcal{A}_{\zeta \mathbf{u}}(\hbar \omega ) \, ,\\
\label{eq:equilibrium_photonenergycurrent}
I^E_{\gamma}&=& \sum_{\zeta} \int d\Omega_{\mathbf{u}} \int d(\hbar \omega)  \hbar \omega \mathcal{D}_{\zeta \mathbf{u}}(\hbar \omega ) \nonumber \\
&& \times \big[N(\hbar \omega)  -  {\cal B}(\hbar \omega) \big] \mathcal{A}_{\zeta \mathbf{u}}(\hbar \omega) \, .
\end{eqnarray}
Expressions~(\ref{eq:equilibrium_photoncurrent}) and (\ref{eq:equilibrium_photonenergycurrent}) are similar to the ones obtained in Ref.~\onlinecite{richter08} dealing with non-equilibrium photon Green's functions.
Interestingly, our approach suggests that in the case of a  non-equilibrium nanosystem given by Eqs.~(\ref{eq:ne_photoncurrent}) and (\ref{eq:ne_photonenergycurrent}),  a generalized energy flow law would involve two-dimensional spectral functions, as $j^E_{\zeta \mathbf{u}}(\epsilon,\hbar \omega)=a_{\zeta \mathbf{u}}(\epsilon,\hbar \omega)[ N_{\zeta \mathbf{u}}(\hbar \omega)-  {\cal B}_{\zeta \mathbf{u}}(\epsilon,\hbar \omega )]$ with $a_{\zeta \mathbf{u}}(\epsilon,\hbar \omega)= [\mathcal{T}^{-}_{\zeta \mathbf{u}}(\epsilon+\hbar \omega,\hbar \omega) - \mathcal{T}^{+}_{\zeta \mathbf{u}}(\epsilon,\hbar \omega )]/h$, following the idea of Ref.~\onlinecite{richter08}.


\section{Entropy current}\label{sec:entropycurrent}

Equation~(\ref{IEphoton}), which gives the energy current from the photon bath to the dot, allows us to calculate the entropy current $I^S$ flowing from the central region to the reservoirs in terms of Green's function and self-energies.

\subsection{Spectral entropy current}\label{sec:spectralentropy}

The device is an open interacting nanosystem connected to three reservoirs: the two electron left and right reservoirs, and the photon bath. In this three-terminal configuration, the entropy current flowing from the central region to the three reservoirs is defined as
\begin{eqnarray} \label{entropyflow}
I^S &=& \frac{ I^h_L }{T_L}+\frac{ I^h_R}{T_R} + \frac{I^h_{\gamma} }{T_{\gamma}} \\
&=&- \frac{\langle \dot H_L\rangle -\mu_L \langle \dot N_L\rangle}{T_L} -\frac{\langle \dot H_R\rangle -\mu_R \langle \dot N_R \rangle}{T_R} \nonumber \\
&&- \frac{\langle \dot H_{\gamma}-\mu_{\gamma} \dot N_{\gamma} \rangle}{T_{\gamma} } \\
&=& \frac{ I^E_L }{T_L}+\frac{ I^E_R}{T_R} - I_L  \Big[ \frac{\mu_L}{T_L} + \frac{\mu_R }{T_R}\Big] + \frac{I^E_{\gamma} }{T_{\gamma}} - I_{\gamma} \frac{ \mu_{\gamma} }{T_{\gamma}}   ~, \nonumber \\
\end{eqnarray}
in which we use the relation $\langle \dot N_L \rangle =-\langle \dot N_R \rangle$ guaranteed by the charge conservation.

Implementing results of Secs.~\ref{photonenergycurrent} and~\ref{energyconservation} in Eq.~(\ref{entropyflow}), we are hence able to derive the entropy current in terms of Green's functions from the spectral entropy current $J^S(\epsilon)$ as follows
 \begin{eqnarray}\label{entropy_current}
I^S&=& \int d\epsilon J^S(\epsilon) ~,
\end{eqnarray}
with 
 \begin{eqnarray}\label{spectral_entropy_current}
J^S(\epsilon)&=& \frac{2}{h}\re \mathrm{Tr} \big[{\bf G}^r{\bf \mathfrak{S}}^<+{\bf G}^<{\bf  \mathfrak{S}}^a\big](\epsilon)~,
\end{eqnarray}
and
\begin{eqnarray}\label{self_entropy}
{\bf  \mathfrak{S}}^{<,a}(\epsilon)&=&\sum_{\alpha \in L,R} \frac{\big(\epsilon -\mu_{\alpha}\big){\bf \Sigma}^{<,a}_{\alpha}(\epsilon)}{T_{\alpha}} \nonumber \\
&&+\frac{1}{2}\frac{{\bf \Xi}^{<,a}_{\gamma}(\epsilon)-\mu_{\gamma} {\bf \widetilde \Sigma}^{<,a}_{\gamma}(\epsilon) }{T_{\gamma}} \, .
\end{eqnarray}

In nanosystems maintained in out-of-equilibrium steady states, the entropy current flowing from the central region to the reservoirs is equal to the rate of entropy production $\Pi =I^S $ \cite{esposito10a}.

\subsection{Entropy production is recast in terms of efficiencies}\label{entropy}

For photovoltaic-thermoelectric converters, we define the nanodevice efficiency as the ratio of the output electrical power or useful heat current to the input power in the form of light, which is given by the heat current of the absorbed photons, $I^h_{abs}$. 
This definition contrasts with ``the maximal power conversion efficiency" defined in practice at maximal output power, and where the denominator is the incident radiant power; it thus does not depend on the processes undergone by the system~\cite{einax11}.

In this section, we focus on three devices based on a central region interacting with light: a photovoltaics ($PV$), a refrigerator based on a cooling by heating process  ($CBH$) \cite{cleuren12}, and finally a joint device which provides both cooling and electrical energy production  ($JCEP$) \cite{entin15}.
For the three nanodevices, the rate of entropy production is recast in terms of efficiencies according to the device, as Whitney proposed in Ref.~\onlinecite{whitney15}. Indeed, all nanodevices ($ND$) provide the same formal rate of entropy production
\begin{eqnarray}\label{formal_P}
\Pi &=&\Pi _{0}\big[ \eta^{rev}_{ND}- \eta_{ND}\big]\, ,
\end{eqnarray}
where $\eta^{rev}_{ND}$ is the efficiency of the reversible nanodevice,  $\eta^{rev}_{ND}I^h_{abs}$ is the output power in the reversible nanodevice, and $\Pi _{0} \eta^{rev}_{ND}$ is the maximum rate of entropy production achievable in the nanodevice. Ratio $\Pi _{0} / \Pi$ reflects how close to the maximum efficiency the device is working.

Table~\ref{efficiencies} summarizes the definitions and notations of the relevant efficiencies discussed for the three nanodevices. These efficiencies are named thermodynamic efficiencies as they can be manipulated following the laws of thermodynamics. 
\begin{table}[htp]
\caption{Efficiency notations and definitions which will be used for a photovoltaic ($PV$), cooling by heating ($CBH$) and a joint cooling and electrical energy production ($JCEP$) nanodevices. For $PV$ and $JCEP$ devices, $I^e_L V\le0$. For a $CBH$ device, $I^h_R \ge 0$, and for a $JCEP$ device, $I^h_L \ge 0$. In all cases, $I^h_{abs} \ge 0$.}
\begin{center}
\begin{tabular}{cc}
\hline \hline
Nanodevice (\textit{ND}) & $\eta_{ND}$   \\ \hline
\begin{tabular}{c} \textit{PV} \\  ($T_L=T_R$, $\mu_L > \mu_R$)  \end{tabular} & $\eta_{PV}=-\frac{I^e_L V}{I^h_{abs}}$  \\
\begin{tabular}{c} \textit{CBH} \\  ($T_L > T_R$, $\mu_L = \mu_R$)  \end{tabular} & $\eta_{CBH}=\frac{I^h_R}{I^h_{abs}}$  \\
\begin{tabular}{c} \textit{JCEP} \\  ($T_L< T_R$, $\mu_L> \mu_R$)  \end{tabular} &  $ \begin{array}{ll} 
\eta_{JCEP}  & =\frac{I^h_L-I^e_LV}{I^h_{abs}} \\ 
& =\eta_{CBH}+\eta_{PV} 
\end{array} $ \\
\multicolumn{2}{c}{}\\
Standard engine  & $\eta^{rev}$ \\ \hline
\begin{tabular}{c}  Carnot machine  (C) \\ ($T_{c(cold)}<T_{h(hot)}$) \end{tabular} & $\eta^{ch}_{\text{C}}=1-\frac{T_c}{T_h}$ \\
Refrigeration ($\circledast$) & $ \eta_{\circledast}^{ch}=\frac{T_c}{T_h-T_c}$ \\
Heat pump (${\wr\wr\wr}$) & $\eta^{ch}_{\wr\wr\wr}=\frac{T_h}{T_h-T_c}$ \\
 \begin{tabular}{c}  Trithermal heat engine ($3T$) \\ ($T_c<T_{i(intermediate)}<T_h$) \end{tabular}&  $\eta_{3T}^{cih}=\eta_{\text{C}}^{ih}\times \eta_{\circledast}^{ci}$ \\
 \hline \hline
\end{tabular}
\end{center}
\label{efficiencies}
\end{table}%

\subsubsection{Standard photovoltaics}

For $T_L=T_R$, we can derive the photovoltaic case
\begin{eqnarray}\label{etaPVND}
\Pi &=&\frac{I^h_{abs}}{T_L}\Big[\underbrace{\Big(1-\Big|\frac{I^h_{em}}{I^h_{abs}}\Big|\Big)\eta^{LS}_{\text{C}}}_{\eta^{rev}_{PV}}-\eta_{PV}\Big] \, ,
\end{eqnarray}
where $I^h_{em}=I^h_{em,st}+I^h_{em,sp}$, and $I^h_{em,st(p)}=I^E_{em,st(p)}+\mu_{\gamma}R_{em,st(p)}$. We always have $I^h_{em}<0$ while $I^h_{abs}>0$. The Carnot efficiency $\eta^{LS}_{\text{C}}$ is defined Tab.~\ref{efficiencies}.

Here, it can be worth deriving the related electroluminescent ($EL$) case, for which $N_\gamma=0$ implies $I^h_{ind}=0$,
\begin{eqnarray}
\Pi &=&I^e_LV \frac{T_{E}-T_L}{T_{E}T_L}\Big[\eta^{LE}_{\wr\wr\wr}-\eta_{EL}\Big]\, ,
\end{eqnarray}
where $\eta_{EL}=-\frac{I^h_{em,sp} }{P^{e}}$ (with $P^{e}=I^e_L V\ge 0$) is the efficiency of the electroluminescent device, and $T_E$ replaces $T_{\gamma}$ is the temperature of the photon bath formed by electroluminescence.

The efficiency of the reversible photovoltaic nanodevice is reduced compared to the Carnot limit: from expression (\ref{etaPVND}), the maximum value of the efficiency is $\eta^{rev}_{PV}=\eta^{LS}_{\text{C}}\Big[1-|\frac{I^h_{em}}{I^h_{abs}}|\Big]$. This maximal efficiency may be compared to the Landsberg's limit~\cite{landsberg98,green}: $\eta_{\text{Landsberg}}= \eta^{AS}_{\text{C}}\Big[1-\frac{1}{3}\frac{T_A}{T_{\gamma}}\big(1+\frac{T_A}{T_{\gamma}}+\frac{T^2_A}{T^2_{\gamma}}\big)\Big]$ ($A$ stands for ambient, and it corresponds to $T_L$ in this work).
Landsberg wanted to reconsider the limit of the Carnot efficiency as the upper limit for photovoltaics. Starting from the model of a dithermal engine, he included the energy and entropy fluxes related to the emission process. In the Landsberg's approach, the central region is a converter in a state of equilibrium, and it behaves as a black body emitting photons at temperature $T_C$ ($C$ stands for converter). Landsberg demonstrated that the maximal efficiency of the reversible device, $\eta_{\text{Landsberg}}$, is reached when $T_C=T_A$. The Landsberg's approach ignores the details of the electron properties in the converter which is also assumed at equilibrium.
However, despite these differences, NEGF-based expression of the entropy production Eq.~(\ref{etaPVND}) comes to  similar conclusions to those of Landsberg:  the maximum efficiency is always lesser than the Carnot limit of a heat engine producing work from the electron and photon reservoirs.

\subsubsection{Cooling by heating process}

We discuss the coefficient of performance of a cooling by heating process as proposed in Ref.~\onlinecite{cleuren12} with $T_L>T_R$ and $V= 0$ (see Tab.~\ref{efficiencies} for the efficiency definitions),
{\small \begin{eqnarray}
\Pi  &=&\frac{I^h_{abs}}{T_L}\Big[\underbrace{\Big(1-\Big|\frac{I^h_{em}}{I^h_{abs}}\Big|\Big)\eta_{3T}^{RLS}}_{\eta^{rev}_{CBH}}-\eta_{CBH} \Big] \, .\nonumber\\
\end{eqnarray}}
From this formula, we deduce for this original cooling process
\begin{eqnarray}
\eta_{CBH}&=&\frac{T_R}{T_L-T_R}\bigg[\Big(1-\Big|\frac{I^h_{em}}{I^h_{abs}}\Big|\Big)\eta_{\text{C}}^{LS}-\frac{T_L \Pi }{I^h_{abs}}\bigg]\, ,
\end{eqnarray}
which meets Eq. (11) of Ref.~\onlinecite{cleuren12} with the additional reducing contribution $|\frac{I^h_{em}}{I^h_{abs}}|\eta^{LE}_{\text{C}}$ to the $CBH$ coefficient of performance. Indeed, the emission processes were not included in the approach of Ref.~\onlinecite{rutten09}, which was developed in the strong optical coupling regime. Moreover,  in the recent model proposed by Wang and co-authors in Ref.~\onlinecite{wang15} to verify the third law of thermodynamics in the refrigerator, the cooling regime includes a parasitic emission in the regime of weak coupling to the electron reservoirs, which involves a single emission wavelength.

\subsubsection{Joint cooling and energy production}

For a more general case, but with a specific device objective, we examine the joint cooling and energy production proposed in Ref.~\onlinecite{entin15}. The joint process can be seen as a photovoltaic configuration with $T_L <T_R$, or a cooling by heating configuration with $V>0$. It follows two expressions for the rate of entropy production
\begin{equation}\label{entropy_JCEP_CBH}
\Pi =I^h_{abs} \frac{T_R-T_L}{T_LT_R} \Big[\eta^{rev}_{CBH}-\eta_{PV}\big(\eta_{\circledast}^{RL}-1\big) -\eta_{JCEP}\Big]\, ,
\end{equation}
and
\begin{equation}\label{entropy_JCEP_PV}
\Pi =\frac{I^h_{abs}}{T_R}\Big[\eta^{rev}_{PV} -\eta_{CBH}\Big(\frac{1-\eta_{\circledast}^{RL}}{\eta_{\circledast}^{RL}}\Big) -\eta_{JCEP}\Big]\, .
\end{equation}
From expressions (\ref{entropy_JCEP_CBH}) and (\ref{entropy_JCEP_PV}), we deduce
\begin{eqnarray}
\eta_{\circledast}^{RL}<1\Rightarrow \eta^{rev}_{CBH}<\eta^{rev}_{JCEP}<\eta^{rev}_{PV} \label{JCEP_CBH} \, ,\\
\eta_{\circledast}^{RL}>1 \Rightarrow  \eta^{rev}_{PV}<\eta^{rev}_{JCEP}<\eta^{rev}_{CBH}\label{JCEP_PV}\, .
\end{eqnarray}
In terms of applications, it means that for $\eta_{\circledast}^{RL}<1$, a joint process more efficiently converts the photon bath power than the $CBH$ one as shown by Entin-Wohlman and co-authors in Ref.~\onlinecite{entin15}, and we outline here that for $\eta_{\circledast}^{RL}>1$, a joint process also more efficiently converts the photon bath power than the \emph{photovoltaic device}. 
Additionally, writing $\eta^{rev}_{JCEP}=\eta^*_{CBH}+\eta^*_{PV}$, expressions (\ref{JCEP_CBH}) and (\ref{JCEP_PV}) also show $\eta^*_{CBH}<\eta^{rev}_{CBH}$ and $\eta^*_{PV}<\eta^{rev}_{PV}$, which means that the single conversion ($CBH$ or $PV$) in the hybrid device is always less efficient than in the corresponding standard device.


\section{Discussing the entropy production in a two-level model including light-matter interaction}\label{sec:negentropy}

In this section, we discuss how the second law of thermodynamics is not automatically verified depending on the model used to simulate how the nanodevice works. We focus on a minimal model of QD-based nanojunction. Such ultimate nanostructures allow us to grasp the essentials of the energy conversion at the nanoscale using three-terminal configurations~\cite{rukola12,sanchez12,jordan13,argawalla16}, which provides the separation between the charge and heat transport, and, at the same time, may motivate innovative experimental realizations, as demonstrated recently in Refs.~\onlinecite{sanchez11} and \onlinecite{thierschmann15}.

\subsection{Basics of the modeling}

\begin{figure}
    \begin{center}
        \includegraphics[width=8 cm]{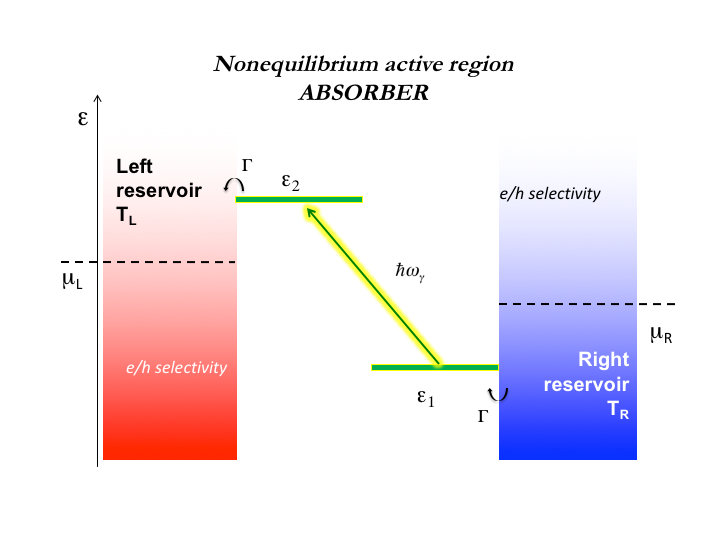}
        \caption{Level diagram of the junction: two
        discrete energy levels are available in the QD at energies
        $\varepsilon_1=0.5$~eV and $\varepsilon_2=-0.5$~eV respectively. Level $1$($2$) is connected to the right(left) electron reservoir only, which provides a perfect electron/hole selectivity. For the numerical calculations, we use $\Gamma=5\,10^{-2}$~eV (half of the imaginary part of the advanced contact self-energy) and $M=10^{-3}$~eV (optical coupling). At the optical resonance, $\hbar \omega_{\gamma}=1$~eV.}
        \label{system_OE}
    \end{center}
\end{figure}

We model a photovoltaic-thermoelectric junction based on quantum dots as shown Fig.~\ref{system_OE}. We follow a simplified methodology within the framework of NEGFs.
The central region is made of a quantum dot giving a nanoscopic interacting volume of 1~nm$^3$. The dot is described by two energy levels which interact with resonant monochromatic radiation $\hbar \omega_{\gamma}$ through the optical coupling $M_{\gamma}$. The upper dot level $\epsilon_2$ is only connected to the left electron reservoir while the lower level $\epsilon_1$ is connected to the right one, which forces the charge separation without applying an electric field. Contact self-energies are given by $ \Sigma^{r,a}_{L,R}(\epsilon)=\Lambda_{L,R}(\epsilon)\mp i\Gamma_{L,R}(\epsilon)/2$~\cite{jauho94}.

The rate of entropy production is calculated in the regime of strong coupling to the reservoirs: calculations are performed at the second order perturbation upon the optical coupling~\cite{berbezier13APL,entin14}. This approach is valid as long as the optical coupling is lower than the transfer parameter, $M_{\gamma} \ll \Gamma_{L,R}$. The bias voltage is symmetrically applied $\mu_{L,R}=\pm eV/2$.

Here we specify the basic case of a linearly polarized monochromatic plane wave as an incident radiation of energy $\hbar\omega_{\gamma}$ and polarization $ \mathbf{\zeta}$.
Functions $\Xi$ and $\widetilde \Sigma$ distinguish between the three radiative processes that are absorption, stimulated emission and spontaneous emission. 
For the monochromatic case, we used 
\begin{subequations} \begin{align}
{\bf \widetilde \Sigma}^{\lessgtr}_{abs}(\epsilon)&=\pm N_{\gamma}  \mathbf{M}_{\gamma} \mathbf{G}^\lessgtr(\epsilon\mp \hbar \omega_{\gamma} ) \mathbf{M}_{\gamma}\label{ximono_a}  \, ,\\
 {\bf \widetilde \Sigma}^{\lessgtr}_{em,st}(\epsilon)&= \mp N_{\gamma}  \mathbf{M}_{\gamma} \mathbf{G}^\lessgtr(\epsilon \pm \hbar \omega_{\gamma} ) \mathbf{M}_{\gamma} \label{ximono_st}  \, , \\
 {\bf \widetilde \Sigma}^{\lessgtr}_{em,sp}(\epsilon)&= \mp \mathbf{M}_{\gamma} \mathbf{G}^\lessgtr(\epsilon \pm \hbar \omega_{\gamma} ) \mathbf{M}_{\gamma} \label{ximono_sp}  \, ,
 \end{align} 
\end{subequations}
with ${\bf \Xi}^{\lessgtr}_{abs\text{ or }em,st}= \hbar\omega_{\gamma} {\bf \widetilde \Sigma}^{\lessgtr}_{abs\text{ or }em,st}$, and the real parts of the retarded and advanced components of the interaction self-energies are ignored.
Optical coupling $\mathbf{M}_{\gamma}$ reads as $M_{nm,\omega_{\gamma} \zeta}=\sqrt{\hbar e^2/2V\epsilon_0\epsilon_r\omega_{\gamma}}P_{nm,\zeta }$ where $P_{nm,\zeta }$ is the momentum matrix element and $V$ the volume of the interacting region.

For the two-level model, the problem is block-diagonal, and we use analytics to derive the spectral currents to the second order in $M_{\gamma}=M_{12,\omega_{\gamma}  \zeta}$ in order to discuss the entropy production in the device. 

\subsection{Particle current}

In this nanojunction design, for which no charge carrier flows without light-matter interaction, we verify that  $\langle \dot N_L \rangle =-\langle \dot N_R \rangle = -\langle \dot N_{\gamma} \rangle$.

We focus on the spectral particle current $J_R(\epsilon)$, $I_R= \int d\epsilon J_R(\epsilon)$, which is positive when an electrical power is produced. To the second order in $M_\gamma$, we find
\begin{eqnarray}\label{eq:JR}
J_{R}(\epsilon)&=&\frac{1}{h}M_{\gamma}^2  \mathcal{A}_1(\epsilon) \mathcal{A}_2(\epsilon+\hbar \omega_{\gamma}) \nonumber \\%
&& \times \Bigg[ N_{\gamma} F^+_{RL}(\epsilon) - (N_{\gamma}+1)F^-_{LR}(\epsilon+\hbar \omega_{\gamma})\Bigg] ~,\nonumber \\
\end{eqnarray}
with
\begin{eqnarray}
\mathcal{A}_1(\epsilon)&=&\frac{\Gamma_R(\epsilon)}{\big(\epsilon- \Lambda_{R}(\epsilon)-\epsilon_1\big)^2+\Gamma_{R}^2(\epsilon)/4 }= -2\im G^r_{11}(\epsilon)~, \nonumber \\ \\
\mathcal{A}_2(\epsilon)&=& \frac{ \Gamma_{L}(\epsilon)}{(\epsilon- \Lambda_{L}(\epsilon)-\epsilon_2)^2+\Gamma_{L}^2(\epsilon)/4 }=-2\im G^r_{22}(\epsilon) ~, \nonumber \\
\end{eqnarray}
and
\begin{eqnarray}
F^{\pm}_{\alpha \beta} (\epsilon)&=&f_\alpha(\epsilon) \big[1-f_\beta(\epsilon\pm\hbar \omega_{\gamma} ) \big]~,
\end{eqnarray}
for $\alpha  \in \{L,R\}$ and $ \beta \in \{L,R\}$. Functions $F^{\pm}_{\alpha \beta} (\epsilon)$ naturally relate the photocurrent to the recent interpretation of the different contributions to the non-symmetrized noise in a quantum dot~\cite{zamoum16}.
From $1-f(\epsilon)=e^{\frac{\epsilon-\mu}{k_B T}} f(\epsilon)$ and $N(\hbar \omega) +1=e^{\frac{\hbar \omega}{k_B T}} N(\hbar \omega)$, $J_R(\epsilon)$ takes  the form
\begin{eqnarray}\label{JR}
J_R(\epsilon) &=& J_{abs}(\epsilon) \Big[ 1-e^{-X(\epsilon)} \Big] ~,
\end{eqnarray}
with 
\begin{eqnarray}
X(\epsilon)&=&\frac{\epsilon + \hbar \omega_{\gamma} -\mu_L}{k_BT_L} -\frac{\epsilon -\mu_R}{k_BT_R} -  \frac{\hbar \omega_{\gamma}}{k_BT_{\gamma}}~,
 \end{eqnarray}
and
\begin{eqnarray}
J_{abs}(\epsilon) &=&\frac{1}{h} M_{\gamma}^2  \mathcal{A}_1(\epsilon) \mathcal{A}_2(\epsilon+\hbar \omega_{\gamma}) \nonumber \\%
&& \times N_{\gamma} f_R(\epsilon)\big(1-f_L(\epsilon+\hbar \omega_{\gamma})\big) \, .
 \end{eqnarray}

For $T_L=T_R$, the $\epsilon$ dependence of $X$ drops out and $X=(\hbar \omega_{\gamma}\eta^{L\gamma}_{\text{C}}-eV)/k_BT_L$, where the Carnot efficiency $\eta^{ch}_{\text{C}}$ is defined Tab.~\ref{efficiencies}. For reservoirs at the same temperature, the spectral particle current, and hence the charge current, vanishes at a voltage called open-circuit voltage in photovoltaics, $eV_{oc}=\hbar \omega_{\gamma}\eta^{L\gamma}_{\text{C}}$ given by $X=0$. This result shows an interesting analogy with the observations made by S\'anchez and B\"{u}ttiker in Ref.~\onlinecite{sanchez12}. Indeed, in the three-terminal configuration they propose, the heat current is controlled from Coulomb interaction instead of the light-matter one. The authors also evidence a stall voltage, Eq.~(16) of Ref.~\onlinecite{sanchez12}, that is an analogue of $V_{oc}$, for which both charge and heat currents vanish. Comparing the two three-terminal configurations, this voltage is a fraction, equal to the Carnot efficiency, of the relevant energy quantum: the charging energy $E_C$ in Ref.~\onlinecite{sanchez12} versus the photon energy $\hbar \omega_{\gamma}$ in the configuration studied here. Moreover, the heat energy current in the right reservoir is also zero at this voltage in the current configuration due to $J^h_R(\epsilon)=(\epsilon-\mu_R) J_R(\epsilon)$. We moreover conclude that the heat current exchanged with the photon reservoir is also zero at $V_{oc}$ using this model, following $I^h_{\gamma}=\hbar \omega_{\gamma}I_{\gamma}=\hbar \omega_{\gamma}I_R$.

\subsection{Entropy production}

Assuming $T_L>T_R$, we calculate the spectral current $J^{\Pi}(\epsilon)$ which gives the rate of entropy production $\Pi $ by integration in the two-level model
\begin{eqnarray}\label{eq:Jpi}
J^{\Pi}(\epsilon)&=&\epsilon J_R(\epsilon) \bigg[ \frac{1}{T_L}- \frac{1}{T_R}\bigg] - J_R(\epsilon) \bigg[ \frac{\mu_L}{T_L} - \frac{\mu_R }{T_R}\bigg]\nonumber \\
&&+\hbar \omega_{\gamma}J_R(\epsilon) \bigg[ \frac{1}{T_L}-\frac{1}{T_{\gamma}} \bigg] \nonumber \\
&=&J_R(\epsilon)\bigg[\frac{\epsilon + \hbar \omega_{\gamma} -\mu_L}{T_L} -\frac{\epsilon -\mu_R}{T_R} -  \frac{\hbar \omega_{\gamma}}{T_{\gamma}}\bigg] ~, \nonumber \\
 \end{eqnarray}
where we used  $\langle \dot N_L \rangle =-\langle \dot N_R \rangle = -\langle \dot N_{\gamma} \rangle$. 
We highlight that this spectral entropy current is not the one obtained from Eqs.~\ref{spectral_entropy_current} and \ref{self_entropy}, but the integrated result gives the same rate of entropy production that in Eq.~\ref{entropy_current}. 

This spectral entropy current depends on $J_R(\epsilon)$, $J_L(\epsilon)$ and $J_\gamma(\epsilon)$.
 We thus obtain
\begin{eqnarray}
\label{spectralentropy}
J^{\Pi}(\epsilon)&=&k_BJ_{abs}(\epsilon)X(\epsilon ) \Big[ 1-e^{-X(\epsilon)} \Big]~.
 \end{eqnarray}
 In Eq.~(\ref{spectralentropy}): on the one hand $J_{abs}$ is always positive as it gives the rate of absorbed photons flowing to the active region, on the other hand the function $X \Big[ 1-e^{-X} \Big]$ is also always positive.
The rate of entropy production is hence always positive, as it results from the direct integration of $J^{\Pi}(\epsilon)$.
When $T_L=T_R$, the entropy current vanishes for $X=0$ like the particle and heat currents, which was discussed in the previous section. Moreover, the analytical model also provides that the entropy production rate is concave at the open-circuit voltage  since we obtain $\partial^2_V I^{\Pi}(V_{oc})=2I_{abs}[\partial_V X(V_{oc})]^2\ge 0$. The rate of the entropy production is minimum for open circuit conditions.

   \begin{figure} [ht]
   \begin{center}
      \begin{tabular}{c} 
   \includegraphics[width=7 cm]{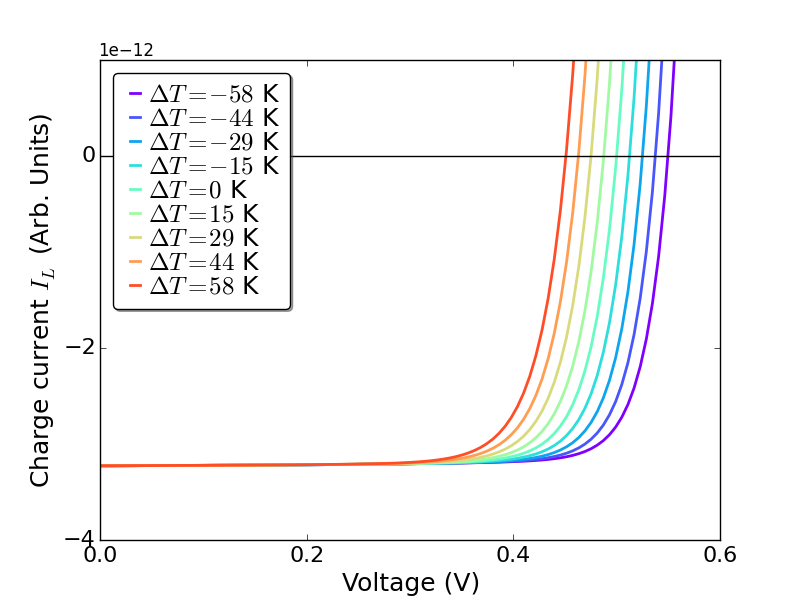} \\
    \includegraphics[width=7 cm]{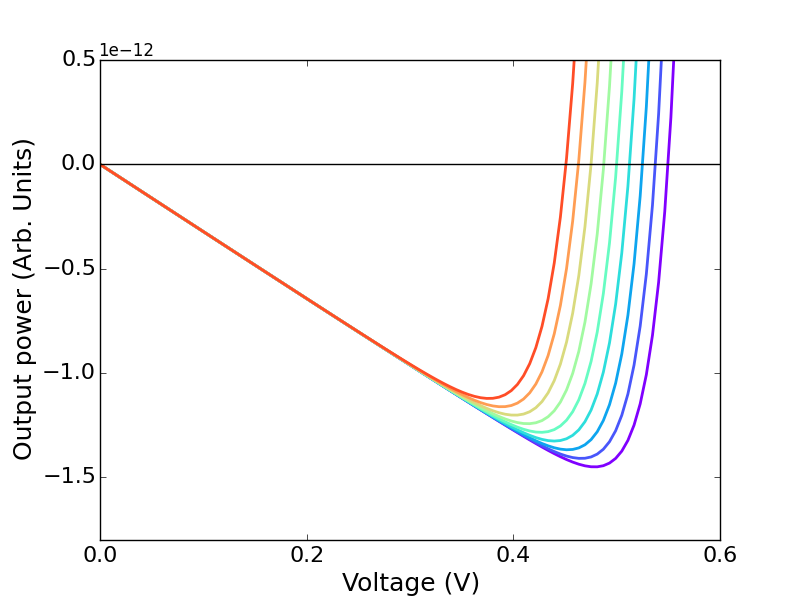} \\
    \includegraphics[width=7 cm]{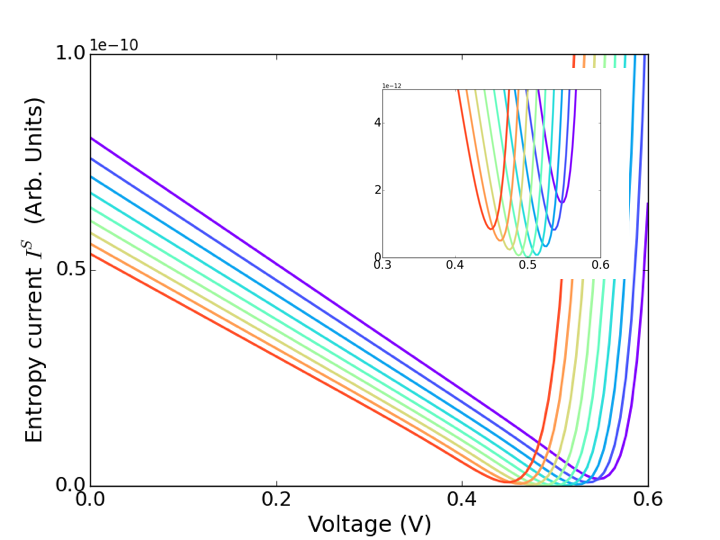}     
       \end{tabular}
   \end{center}
   \caption 
   { \label{fig:2L_ISV} $I_L-V$ and $I^S -V$ curves in the QD-based nanojunction for different temperature gradients $\Delta T=T_R-T_L$, $k_BT_L=0.025$~eV and $k_BT_{\gamma}=0.05$~eV.}
   \end{figure} 

For various temperature gradients in between the two electronic reservoirs, we numerically calculated the charge and entropy currents in the optoelectronic junction depicted Fig.~\ref{system_OE}.  In this numerical implementation, contact self-energies are given by a unique parameter $\Gamma$ in the wide band limit $ \im \Sigma_{L,R}=-i\Gamma/2$ ($\Lambda=0$)~\cite{jauho94}, and the other parameters are given Fig.~\ref{system_OE}.
Characteristics of Fig.~\ref{fig:2L_ISV} show that cooling the right reservoir (cathode) in this configuration enhances the open-circuit voltage $V_{oc}$ and the maximal output power. Additionally, the numerical results confirm that  the rate of entropy production is always positive for all temperature differences and vanishes at $V_{oc}$ for $T_L=T_R$, for which $X$ does not depend on $\epsilon$. Moreover, the numerical results show that the rate is minimum at $V_{oc}$ and that the rate curve is concave around this point.

\subsection{Limits of traditional models}

In optoelectronics, traditional models sometimes modify the presented calculations in the framework of NEGFs to determine the functioning of the device: the number of photons is not given by a Bose-Einstein function but by the incoming photon flux~\cite{wurfel} and the spontaneous emission is integrated over all possible photon states~\cite{SteigerThesis}. 

\noindent \underline{Photon number}. 
Instead of the Bose function taken at 5000K, realistic modelings of solar cells introduce
\begin{equation}
N_\phi=\frac{I_\phi V }{\hbar \omega_\gamma C_0 / \kappa }\, .
\label{eq:Nphi} 
 \end{equation}
Number $N_\phi$ is calculated from the photon flux intensity at the surface of the earth ${I_\phi} \approx 10^3$~Wm$^{-2}$, which accounts for the solid angle between the sun and the earth, the volume and the refractive index, $V$ and $\kappa$, of the absorber (here the nanoscopic dot) and the speed of light in the vacuum $C_0$.

\begin{figure} [ht]
   \begin{center}
  \includegraphics[width=7 cm]{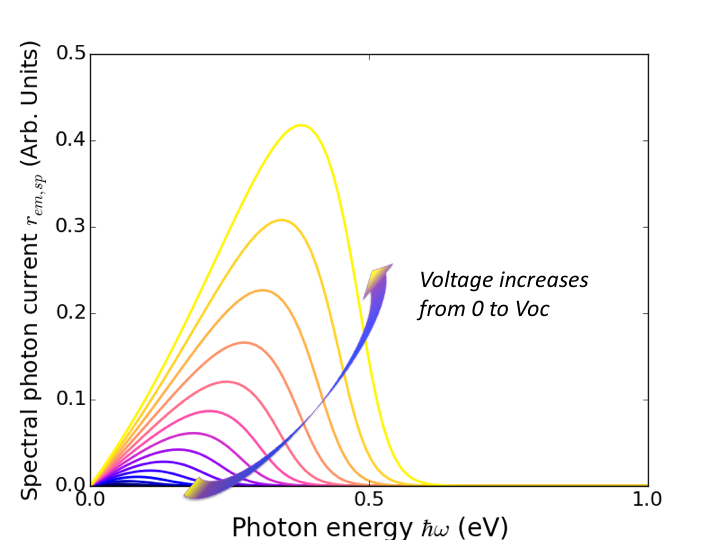} 
    \end{center}
   \caption 
   { \label{fig:emission} Spectral emission current in the QD-based nanojunction following Eq.~(\ref{luminance}) for equal temperatures of electron reservoirs, varying the voltage from zero to $0.5$~V.}
   \end{figure}

\noindent  \underline{Integrated spontaneous emission}.
In contrast with the two induced radiative processes, the spontaneous emission does not explicitly depend on the properties of the incident radiation (but implicitly through the self-consistently determined electron Green's function $G$). We will integrate over all the transition energies available in the interacting region~\cite{SteigerThesis}. 
For nanosystems strongly hybridized with electron reservoirs, the density of states broadens, and a large interval of emission energies is possible, as illustrated Fig.~\ref{fig:emission}, for the set of parameters used Fig.~\ref{system_OE}. Assuming a polarization isotropy for the interacting nanosystem, we introduce the photon density of states $\rho(\hbar \omega)= V (\hbar \omega)^2 \kappa /\pi^2 \hbar^3 C_0^3$ to reformulate
\begin{subequations}\label{eq:integratedemission} \begin{align}
 {\bf \widetilde \Sigma}^{\lessgtr}_{em,sp}(\epsilon)=&\mp \int d(\hbar \omega) \rho(\hbar \omega ) \nonumber \\
 & \times \mathbf{M}(\hbar \omega) \mathbf{G}^\lessgtr(\epsilon\pm \hbar \omega )\mathbf{M}(\hbar \omega) \, ,\\
{\bf \Xi}^{\lessgtr}_{em,sp}(\epsilon)=&\mp \int d(\hbar \omega) \rho(\hbar \omega ) \hbar \omega \nonumber \\
 & \times \mathbf{M}(\hbar \omega) \mathbf{G}^\lessgtr(\epsilon\pm \hbar \omega )\mathbf{M}(\hbar \omega)  \, .
\end{align} 
\end{subequations}
%

\noindent  \underline{Limits}. We here compare three models: A] the Bose function is used and a single emission energy is retained (see Eqs.~(\ref{eq:JR}) and~(\ref{eq:Jpi})); B] the photon occupation number is replaced by the photon flux  (see Eq.~(\ref{eq:Nphi})), and a single emission energy is retained; and C] the Bose function is used and the spontaneous emission is integrated over all photon energies (see Eq.~(\ref{eq:integratedemission})).
Within models B  and C, the energy conservation law still holds but the condition of a thermal photon reservoir fails: $N_\gamma +1 \ne N_\gamma e^{\frac{\hbar \omega_\gamma}{k_BT_{\gamma}}}$, and a positive entropy production is no longer guaranteed. 

\begin{figure} [ht]
   \begin{center}
      \begin{tabular}{cc} 
   \includegraphics[height=3.3 cm]{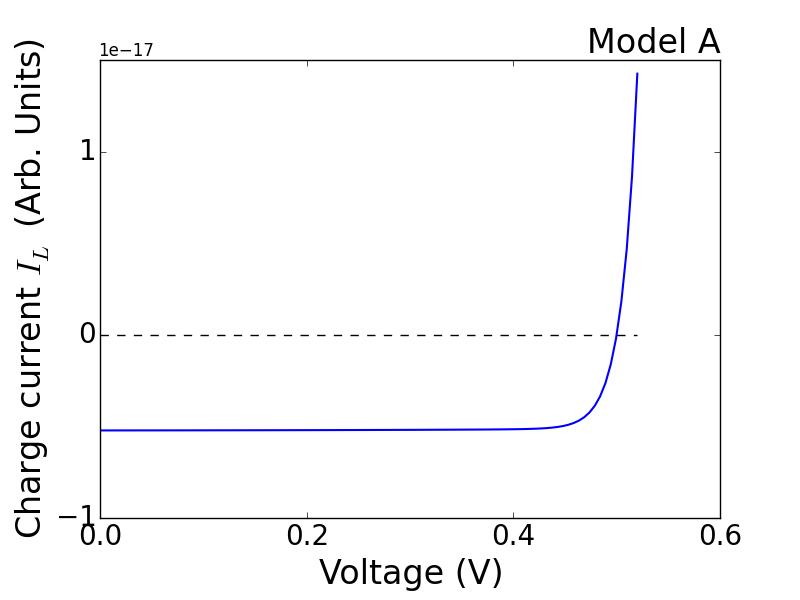} & \includegraphics[height=3.3 cm]{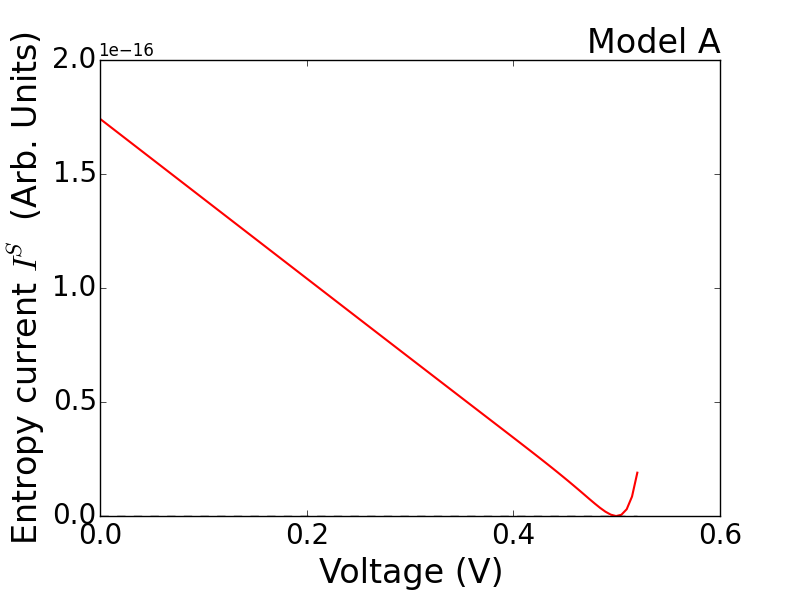} \\
   \includegraphics[height=3.3 cm]{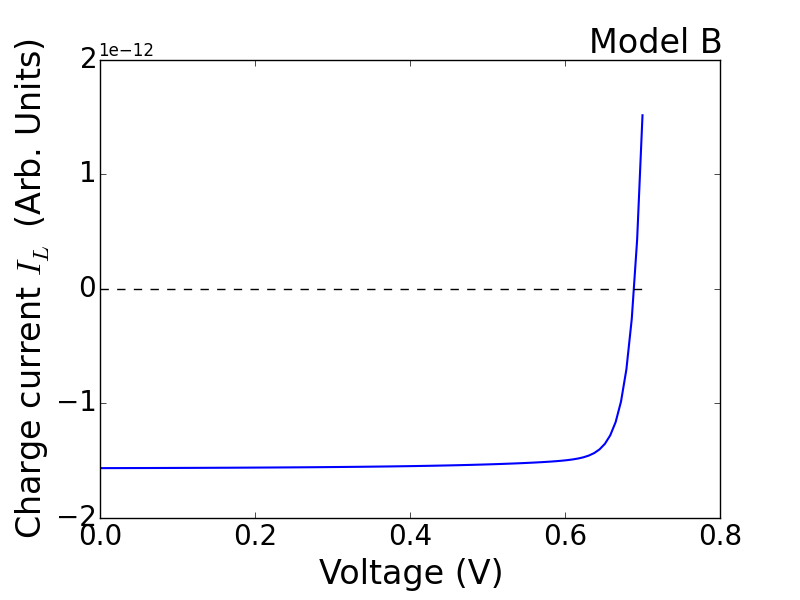} & \includegraphics[height=3.3 cm]{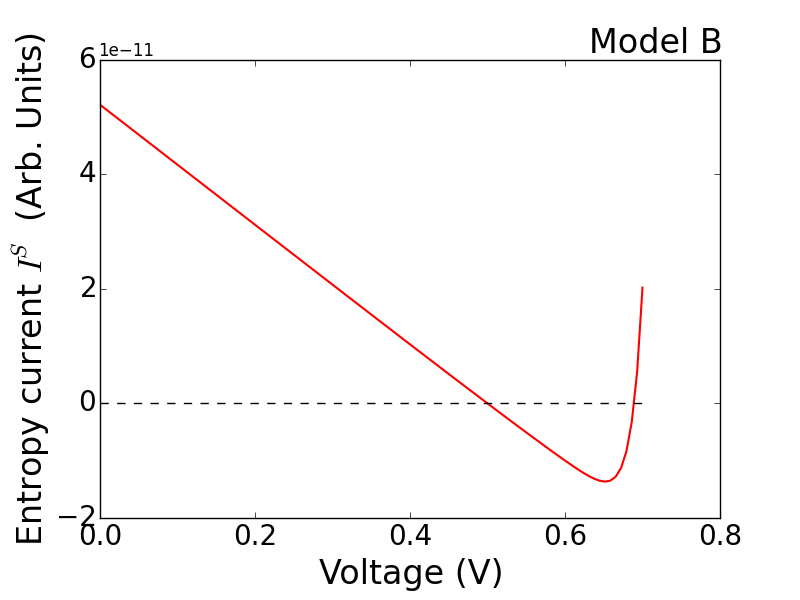} \\
   \includegraphics[height=3.3 cm]{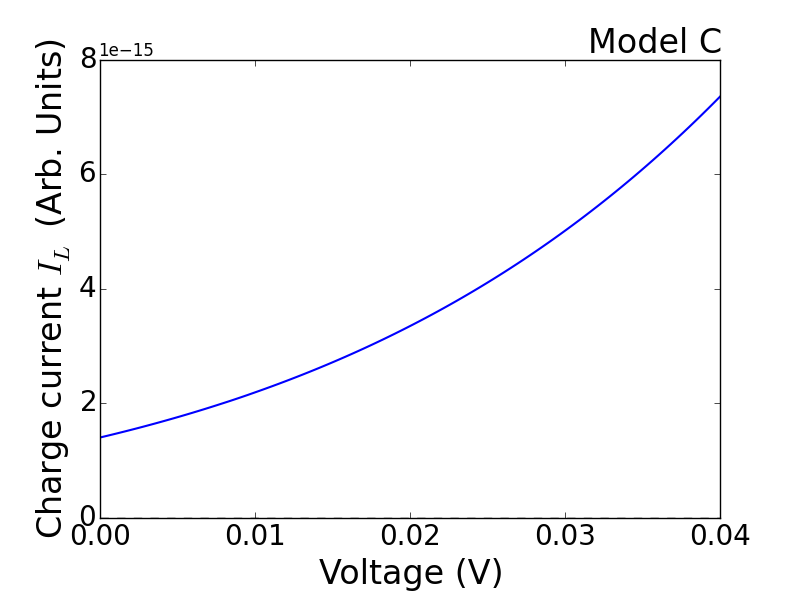} & \includegraphics[height=3.3 cm]{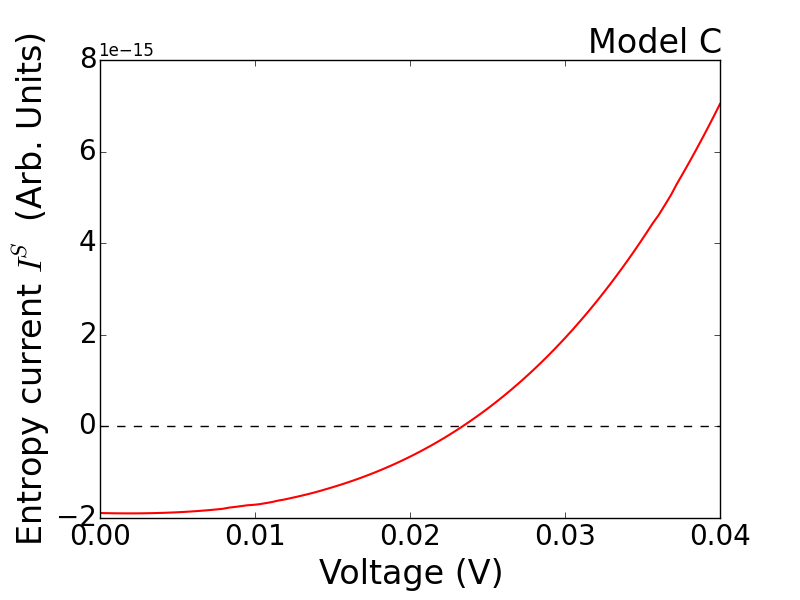}
     \end{tabular}
   \end{center}
   \caption 
   { \label{fig:limits} $I_L-V$ and $I^S -V$ curves in the QD-based nanojunction for the three models A, B and C detailed in the text.}
   \end{figure}

To see this, we numerically implement the three models to determine the charge and entropy currents shown in Fig.~\ref{fig:limits}, with $k_BT_{\gamma}=0.03$~eV and $k_BT_L=0.015$~eV.   In this range of parameters, we observe several dramatic differences.
In this range of parameters, integrating the spontaneous emission contribution over all photon energies leads to a broken photovoltaic nanodevice: with model C, the electrical power is positive for positive bias, which means that the nanodevice is investing electrical power, due to an intense emission current, while a standard photovoltaic nanodevice is expected to produce electrical power until the open circuit voltage is reached, as obtained with models A and B.
Integrating the spontaneous emission contribution over all photon energies brings NEGF formalism to a paradox: on the one hand it is crucial to integrate the spontaneous emission over all photon energies to describe how the device works; on the other hand, we obtain that this integration, which is done after the derivation presented in Secs.~\ref{sec:hamiltonian} and \ref{sec:currents}, provides a whole model which does not verify the second law of thermodynamics.
 
Different authors showed that entropy production is positive in the scope of Landauer or NEGF formalisms~\cite{whitney15,esposito15,yamamoto15}, but all of their works concerned non-interacting systems. In interacting systems with proportional coupling to reservoirs, it seems possible to show via a Landauer-\textit{like} formulation~\cite{meir92} that entropy production is positive. But photovoltaics is very far from this condition of proportionality, since the electron/hole selectivity requires very different couplings to the left and right reservoirs. 
In the current discussion, the light-matter interaction is included to model a photovoltaic device. In the case of a two-level system, we have demonstrated with the help of analytical calculations that entropy production is positive when one assumes a thermal photon reservoir and keeps the same photon energy as the  photon source.
We show that deviation from these two conditions no longer provides a model that guarantees the second law of thermodynamics, inside the framework exposed here.
Model-dependent violations of the second law have already been discussed in the field of photovoltaics~\cite{luque97}. 
Actually, it could seem surprising that an empirical model complies with the second law of thermodynamics for all range of model parameters. In the two highlighted models, B and C, modifications are done after the straightforward derivation of the currents beginning from the hamiltonian model. Models B and C can be thus regarded as empirical models.
Indeed, models B as C break the fluctuation-dissipation relation between the photon Green's functions~\cite{bruus}, which is related to the time-reversal symmetry, and, hence, micro-reversibility~\cite{campisi11}.


\section{CONCLUSION}

Using non-equilibrium electron Green's functions,  we derived the formal expressions of photon energy and particle currents inside a non-equilibrium nanodevice only interacting with light. 
Conducted to their quasi-equilibrium limits, these expressions refresh the issue of local quasi-equilibrium parameters, which provides informations about the absorbing region from the analysis of re-radiated light. Temperatures of out-of-equilibrium populations of electrons and holes could be traced back from the NEGF-based approach used in this paper.
In their general forms, these expressions allow us to formulate the spectral entropy current flowing in the three reservoirs of the nanodevice, namely the two electronic reservoirs, and the photon bath.
Moreover, we recast the net entropy flow in the difference between the efficiency of the reversible photovoltaic-thermoelectric hybrid, and the effective device efficiency. 
Considering quantum dot based hybrid systems, we showed from analytics that entropy production is always positive when a single photon energy is taken for the spontaneous emission process. But, we found from numerics that the rate of entropy production can be negative when more traditional models are followed, in particular when the emission contribution is integrated over all the photon energies of the broadened spectrum. 
This work brings to light this striking paradox and opens avenues to explore entropy production at the nanoscale in interacting systems, which reiterates current fundamental questions underlying the second law of thermodynamics in quantum systems. 

\acknowledgments
We thank R.~Whitney, A. M. Dar\'e, F.~Gibelli, J.-F.~Guillemoles, and L.~Lombez for their interest in this work and for 
valuable discussions on different parts of this work.
We acknowledge financial support from the CNRS Cellule Energie funding project ``ICARE''.


\end{document}